\documentclass{emulateapj}
\shorttitle{Secondary Photons from Hypernovae}

\shortauthors{Asano \& M\'esz\'aros}

\def\siml{\lower4pt \hbox{$\buildrel < \over \sim$}}
\def\simg{\lower4pt \hbox{$\buildrel > \over \sim$}}

\begin{document}

\title{
Secondary Photons from High-energy Protons Accelerated in Hypernovae
}

\author{\scshape Katsuaki Asano\altaffilmark{1} and Peter 
M\'esz\'aros\altaffilmark{2,3,4}}
\email{asano@phys.titech.ac.jp, nnp@astro.psu.edu}

%\date{\today}
%\date{Submitted; accepted}

\altaffiltext{1}{Interactive Research Center for Science,
Tokyo Institute of Technology,
2-12-1 Ookayama, Meguro-ku, Tokyo 152-8550, Japan}
\altaffiltext{2}{Department of Astronomy \& Astrophysics, Pennsylvania State 
University,
University Park, PA 16802}
\altaffiltext{3}{Department of Physics, Pennsylvania State University,
University Park, PA 16802}
\altaffiltext{4}{Center for Particle Astrophysics, Pennsylvania State University,
University Park, PA 16802}

\begin{abstract}

Recent observations show that hypernovae may deposit some fraction of 
their kinetic energy in mildly relativistic ejecta.  In the dissipation 
process of such ejecta in a stellar wind, cosmic ray protons can be
accelerated up to $\sim 10^{19}$ eV.
We discuss the TeV to MeV gamma-ray
and the X-ray photon signatures of cosmic rays accelerated in hypernovae.
Secondary X-ray photons,
emitted by electron-positron pairs produced via
cascade processes due to high-energy protons,
are the most promising targets for X-ray telescopes.
Synchrotron photons emitted by protons can appear in the GeV band,
requiring nearby ($<40$ Mpc) hypernovae for detection 
with GLAST.  In addition, air Cherenkov telescopes may be able to detect
regenerated TeV photons emitted by electron-positron pairs
generated by CMB attenuation of $\pi^0$ decay photons.

\end{abstract}


\maketitle

\section{Introduction}
\label{sec:intro}

Hypernovae (HNe) are a peculiar type of supernova with ejecta velocities and 
apparent isotropic-equivalent ejecta energies which are larger than usual, 
and with indications of anisotropy (Nomoto, et al, 2008). Some of 
them are associated with long gamma-ray bursts (e.g. Woosley and Bloom, 
2006; a recent example being GRB060218/SN2006aj, e.g. Campana et al, 2006),
while others appear not to be.
Cosmic rays up to energies $E\siml 10^{17}$
eV are thought to be accelerated in relatively normal supernova remnants 
(Hillas, 2005).  More recently, it has been suggested that
HNe may 
accelerate cosmic ray protons or nuclei up to energies $E\lesssim 4\times 
10^{18} Z$ eV \citep{wan07,bud07,wan08}.
The smoking gun proof
for cosmic ray acceleration in supernova remnants would be the 
observation of very high energy ($\simg$ TeV) neutrinos (e.g. Kistler and 
Beacom, 2006), which will require completion of cubic kilometer detectors, or 
else the observation of secondary photons arising from pions, which remains 
inconclusive (e.g. Gabici and Aharonian, 2007; Katz and Waxman, 2008; etc.).
Nonetheless, continued gamma-ray observation with air Cherenkov telescopes
(ACTs) in the TeV range, and with GLAST and AGILE in the GeV range may provide
the best immediate hopes for resolving this question. In the present paper,
we consider the same question in relation to HNe, and address the 
question of the photons signatures from secondaries arising from cosmic ray 
acceleration in typical HNe.

In \S \ref{sec:model}, we describe our model NHs and consider
the baryonic and photonic environment in which the explosion occurs, as
well as its effect on the  cosmic rays accelerated in the ejecta.
In \S \ref{sec:results} we discuss the Monte Carlo simulations performed 
on these models, and present the results for the photon signatures arising 
from various secondary components.  In \S \ref{sec:disc} we discuss the 
detectability prospects for these signals, compared to the sensitivity of
ACTs and GLAST, and summarize our results and conclusions.

\section{Model Description}

\label{sec:model}

HNe, especially the Ic types associated with GRB but also
some of the unassociated ones, are thought to be due to WR progenitors
(Nomoto et al, 2006), and as such are expected to have had a strong
stellar wind phase prior to the explosion. The model of a HN 
ejecta expanding in a stellar wind used in this paper is based on the 
model of \citet{wan07}.  Thus, we consider a stellar wind environment
around the progenitor which is characteristic of WR stars.  Assuming a 
mass loss rate $\dot{M}$ and a wind velocity $v_{\rm w}$, the density 
profile of the wind is written as
%%%%%%%%%%%%%%%%%%%%%%%%%%%%%%%%%%%
%\begin{equation}
$\rho(r)=
%\frac{\dot{M}}{4 \pi v_{\rm w}} r^{-2}
5 \times 10^{11} A_{*} r^{-2}$ g $\mbox{cm}^{-1}$,
%\end{equation}
%%%%%%%%%%%%%%%%%%%%%%%%%%%%%%%%%%%
where $A_{*}$ ($\propto \dot{M}/v_{\rm w}$)
$=1$ corresponds to $\dot{M}=10^{-5} M_{\sun} \mbox{yr}^{-1}$
and $v_{\rm w}=10^3 \mbox{km}~\mbox{s}^{-1}$ \citep{wan08}.

The outer envelope regions (the ejecta) of the exploding HN, as shown 
by \citet{sod06}, have a kinetic energy distribution
$\propto (\Gamma \beta)^{-\alpha}$. In this paper we assume
%%%%%%%%%%%%%%%%%%%%%%%%%%%%%%%%%%%
%\begin{equation}
$E_{\rm k}=10^{52} (\Gamma \beta/0.1 )^{-2}$ erg,
%\end{equation}
%%%%%%%%%%%%%%%%%%%%%%%%%%%%%%%%%%%
where the velocity of the bulk of the ejecta ranges from $\Gamma \beta 
\simeq 0.1$ up to semi-relativistic values ($\Gamma \beta \simeq 1$),
where $\beta=v/c$ and $\Gamma$ are the ejecta normalized velocity
and bulk Lorentz factor, respectively.

For non-relativistic ejecta
($\Gamma \beta \lesssim 0.5$, $E_{\rm k} \gtrsim 4 \times 10^{50}$ erg),
the free expansion phase before deceleration sets
in lasts for
%a time
%%%%%%%%%%%%%%%%%%%%%%%%%%%%%%%%%%%
%\begin{equation}
%t_{\rm d,NR} =
$440 ( \Gamma \beta/0.5 )^{-5} A_{*}^{-1}$
days.
%\end{equation}
%%%%%%%%%%%%%%%%%%%%%%%%%%%%%%%%%%%
Therefore, non-relativistic ejecta cannot dissipate their kinetic energy
within the 10-20 days typical timescale of the UV-optical photon radiation
from HNe.  Thus after the optical emission from HNe has declined,
we may not expect secondary photons originating from $p \gamma$-interactions, 
even if a sufficient amount of high-energy protons are produced.

On the other hand, the bulk of the kinetic energy of the mildly relativistic 
ejecta are dissipated at a radius
%%%%%%%%%%%%%%%%%%%%%%%%%%%%%%%%%%%
%\begin{equation}
$R_{\rm d} \simeq 10^{16} ( \Gamma \beta/1.0 )^{-1} A_{*}^{-1}$
cm.
%\end{equation}
%%%%%%%%%%%%%%%%%%%%%%%%%%%%%%%%%%%
Since $\beta \simeq 1$ for such ejecta, all the mildly relativistic ejecta 
dissipate their kinetic energy within 10 days, as long as $A_{*} \gtrsim 1$.
Therefore, hereafter we consider only the most energetic component $\Gamma \beta=1$
($E_{\rm k}=10^{50}$ erg) of the mildly relativistic ejecta.

For the ejecta with $\Gamma \beta=1$, the magnetic field at the dissipation 
radius $R_{\rm d}$ may be estimated as $B^2/8 \pi=\epsilon_B \rho(R_{\rm d}) c^2$,
where $\epsilon_B=0.1 \epsilon_{B,-1}$ is the fraction of
the equipartition value of the magnetic field. Thus, we obtain
%%%%%%%%%%%%%%%%%%%%%%%%%%%%%%%%%%%
%\begin{equation}
$B \simeq 3.4 \epsilon_{B,-1}^{1/2} A_{*}^{3/2}$ G.
%\end{equation}
%%%%%%%%%%%%%%%%%%%%%%%%%%%%%%%%%%%

The maximum energy of accelerated protons for the mildly relativistic
ejecta may be written as $\varepsilon_{\rm max} \simeq e B R$.
With the above values of $B$ and $R_{\rm d}$, we obtain
%%%%%%%%%%%%%%%%%%%%%%%%%%%%%%%%%%%
%\begin{equation}
$\varepsilon_{\rm max} \simeq
10^{19} \epsilon_{B,-1}^{1/2} A_{*}^{1/2}$ eV,
%\end{equation}
%%%%%%%%%%%%%%%%%%%%%%%%%%%%%%%%%%%
for the most energetic ejecta dissipated within 10 days.
Adopting this value, the energy distribution of accelerated protons
is assumed to be $N_p(\varepsilon_p) \propto \varepsilon_p^{-2}
\exp{(-\varepsilon_p/\varepsilon_{\rm max})}$.
The amount of protons is normalized by an efficiency factor $1/6$ 
for the conversion of dissipated kinetic energy ($10^{50}$ erg) into 
energy of accelerated protons \citep{hil05}.

The usually observed low energy ($\sim 1$ eV) photons from HNe are 
attributed
largely to radioactive decay in the non-relativistic ($\Gamma \beta \simeq 0.1$) 
ejecta. After an initial rise and just after the dissipation of the ejecta 
($\sim$ a few days), the changes in the optical luminosity of the HN 
are not very drastic (1-2 magnitudes), as shown in \citet{pia06}.
Therefore, a constant luminosity for a few tens of days may be a reasonable
approximation.  As shown in \citet{maz06}, the peak flux is at 4000-5500 \AA, which 
is considered to be due to broad emission lines of several types of metals.
Here we mimic this photon field by a thermal photon field with a temperature 
$T=1$ eV and luminosity $L_{\rm HN}=10^{43} \mbox{erg}~\mbox{s}^{-1}$,
which is compatible with the bolometric luminosity of SN1998bw
around its peak brightness \citep{gal98}.

Another source of low energy photons is synchrotron radiation from 
electrons accelerated in the ejecta.
For mildly relativistic shocks, the typical Lorentz factor of
accelerated electrons is less than $m_{\rm p}/m_{\rm e} \sim 10^3$,
which implies that the typical energy of synchrotron photons 
is $\varepsilon_\gamma \lesssim 10^{-2}$ eV for $B=3$ G.
The cooling timescale of such electrons is $\sim 1$ day,
which is much shorter than the timescales we consider.
Injection of accelerated electrons may continue
for a few tens of days as indicated by radio observations.
The most luminous radio afterglow in HNe observed ever
is in SN 2003dh \citep{ber03} of
$\sim 10^{41}$ erg $\mbox{s}^{-1}$,
which may slightly enhance pion production efficiency
for the highest energy protons ($>10^{18}$ eV).
Here, for simplicity,
we neglect the photon emission from accelerated electrons.

In order to calculate the pion production and the subsequent
cascade processes, we carry out Monte Carlo simulations using a code
developed for and discussed in a series of studies of gamma-ray burst 
physics \citep{asa05,asa06,asa07}.  We use a simple one-zone approximation,
and follow during a finite time the physical processes of pion production,
pion decay, muon decay, and $\gamma \gamma$ electron-positron pair creation, 
as well as the usual radiation processes of synchrotron (SY) and inverse
Compton (IC) emission from protons, pions, muons, electrons, and positrons.
We assume that the accelerated protons are injected promptly
just after the energy dissipation occurs at $r=R_{\rm d}$.
Since the bulk motion of the post-shock region is non-relativistic,
we neglect the expansion of the ejecta after that.

\section{Results}

\label{sec:results}

\begin{figure}[h]
\centering
\epsscale{1.0}
\plotone{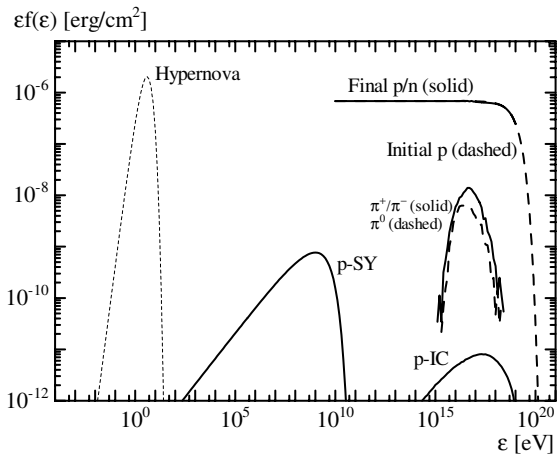}
\caption{
Spectra of the initial protons, and of the surviving protons/neutrons after 
the timescale $R_{\rm d}/c = 3.3 \times 10^5$ s $\sim$ 4 days, for a 
hypernova explosion in a wind characterized by the mass loss rate $A_{*}=1$.
The spectra of secondary particles and photons created during this time scale are 
also plotted, including pions, proton synchrotron photons (labeled p-SY), and 
proton inverse Compton photons (labeled p-IC), not all of which can be directely 
observed. The fluence is normalized assuming a distance $D=100$ Mpc.
The hypernova soft photon spectrum is plotted with thin dashed lines.
\label{fig:cr-a1}}
\end{figure}

First, we consider a HN exploding in a standard wind with $A_{*}=1$.
As shown in Fig. \ref{fig:cr-a1}, even after the accelerated protons
have been irradiated by the HN photons for $R_{\rm d}/c \sim 4$ days,
most of the protons still retain a large fraction of  their energy.
The pion-production timescale $t_{\rm p\gamma}$ is inversely proportional 
to the HN photon density $\propto L T^{-1} R_{\rm d}^{-2}$. Following 
\citet{wan07}, the fraction of the energy lost by protons to pions is estimated
as $f_{\rm p\gamma}\equiv R_{\rm d}/(c t_{\rm p\gamma})=0.04 A_*$, which is 
consistent with our numerical result.
Charged pions of typical energy $\sim 10^{16}$ eV and a total 
fluence of $4.5 \times 10^{46}$ erg are produced within this timescale,
which give rise to electromagnetic cascades.

%\vspace*{2cm}

\begin{figure}[h]
\centering
\epsscale{1.0}
\plotone{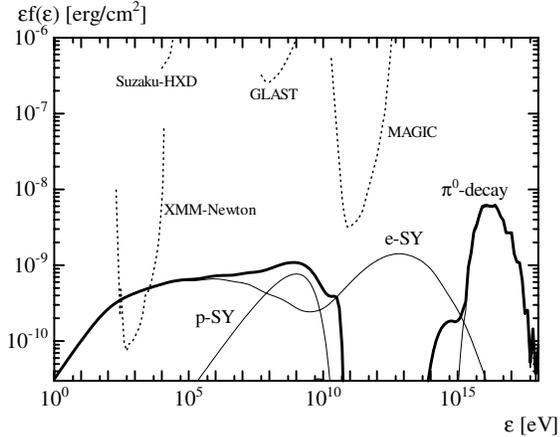}
\caption{
Spectral fluence of combined secondary and tertiary photons from the hypernova 
of Fig. \ref{fig:cr-a1} (thick solid line), at the distance $D=100$ Mpc 
for $A_{*}=1$, integrated over $R_{\rm d}/c = 3.3 \times 10^5$ s. Thin curves 
denote electron/positron synchrotron (labeled e-SY), proton synchrotron
(labeled p-SY), and $\pi^0$-decay components
emitted inside the source.
The resulting photon spectrum escaping from the source, taking into
account internal $\gamma\gamma$ absorption by the hypernova soft photon field,
is shown by heavier lines. The sensitivity curves for representative
instruments are plotted with dotted lines.  The curves for XMM-Newton and 
Suzaku are normalized by 100 ks integration time.
\label{fig:phot-a1}}
\end{figure}

As shown in Fig. \ref{fig:phot-a1}, if one neglects the 
$\gamma \gamma$-absorption effects, a spectral bump of SY
emission appears around the TeV-10 TeV energy range.
This bump is due to SY photons of $15 \epsilon_{B,-1}^{1/2} A_*^{3/2}$
TeV from positrons and electrons from muon decay.
Since the SY photon energy scales as the square of the energy of 
the emitters, the photon bump energy range spreads over 2-3 orders of magnitude.
The inclusion of $\gamma\gamma$ effects leads to these photons 
being absorbed through interaction with the HN soft photon field,
creating electron-positron pairs of 1-10 $\epsilon_{B,-1}^{1/2} A_*^{3/2}$ TeV.
The secondary pairs emit SY photons over a wide energy range
(which should be twice as wide as that of the TeV bump),
from a few keV to a few hundreds of keV.
These pairs cool so promptly that the result is a simple power-law 
spectrum due to cooled pairs below 100 eV.
The secondary photon emission is expected to last as long as
the HN emits optical photons, even though the cooling time of 
the electrons that emit X-ray photons is only about 1 minute.
These photons will be observed with various present-day
X-ray telescopes for $100$ ks integration,
as shown in Fig. \ref{fig:phot-a1}, unless the X-ray afterglow emission 
of a GRB overwhelms it, e.g.  as seen in GRB060218/SN2006aj.

%\vspace*{2cm}

Around the GeV region, proton SY emission yields a bump
in the photon spectrum, as seen in Fig. \ref{fig:phot-a1}.
The typical energy of proton SY photons
is $2.4 \epsilon_{B,-1}^{3/2} A_*^{5/2}$ GeV, 
and the ratio $t_{\rm p \gamma}/t_{\rm p,syn}$
(where $t_{\rm p,syn}=$ SY cooling timescale)
is $\sim 0.2 (A_* \epsilon_{B,-1})^{3/2}$, which roughly agrees 
with the obtained energy fraction of proton SY
to pions. For this distance the  fluence is much lower 
than the GLAST detection limit. To detect 100 MeV photons with GLAST 
would require a HN at $\leq 6$ Mpc.

Another notable feature of Fig. \ref{fig:phot-a1}, around energies 
$\sim 10^{16}$ eV, is the prominent presence of photons from $\pi^0$-decay.
Those photons escape without being absorbed by the $\sim$ 1 eV thermal 
photon field assumed for the HN. However, the mean free path of 
these $10^{16}$ eV photons against $\gamma \gamma$ absorption by 
cosmic microwave background (CMB) photons is $\sim 10$ kpc \citep{aha02}, so 
that we cannot expect to detect such photons directly.
The secondary electron-positron pairs generated by attenuation are very 
energetic, and are inverse-Compton scattered by CMB photons, e.g. as 
discussed for GRBs \citep[and references therein]{raz04,mur07}.
These boosted photons can pair-produce again, and the process repeats itself
until the energy of the degraded photons is in the 1-10 TeV range. The mean
free path of these regenerated 1-10 TeV photons is longer than 100 Mpc,
and they can reach the Earth. As long as the intergalactic magnetic field is 
weak enough, the delay time of TeV photons emitted by $\sim 100$ TeV 
electrons/positrons is negligible in comparison with the timescale a few 
days \citep{raz04}.
We omit plotting the spectrum of the regenerated photons in Fig.
\ref{fig:phot-a1}, since from $D=100$ Mpc it will be hard to detect them;
however, if a HN occurs in the 
Virgo Cluster ($D \sim 20$ Mpc), there would be a chance to detect these 
secondary TeV photons (see below).

Next, we consider a HN occurring in a denser wind with $A_{*}=5$. 
This value is compatible with currently available data on wind mass 
losses suggesting ${\dot M}\lesssim \hbox{few}\times 10^{-4}M_\odot/{\rm yr}$,
which refer to stars well before any explosion \citep{Maeder07}. Physically,
even larger values may be plausible, since one expects the the mass loss 
to increase considerably as the evolution of the core rapidly approaches the 
final collapse, with a rapid increase in the luminosity and the envelope expansion 
rate.

In this case of $A_{*}=5$, the ejecta will stall at $R_{\rm d}
\sim 2 \times 10^{15}$ cm within $\sim 1$ day,
so that the non-relativistic ejecta of $\Gamma \beta=0.1$
can catch up with the decelerating ejecta about 10 days later.
However, at least until the non-relativistic ejecta has
caught up, the secondary photons are largely observable.

\begin{figure}[h]
\centering
\epsscale{1.0}
\plotone{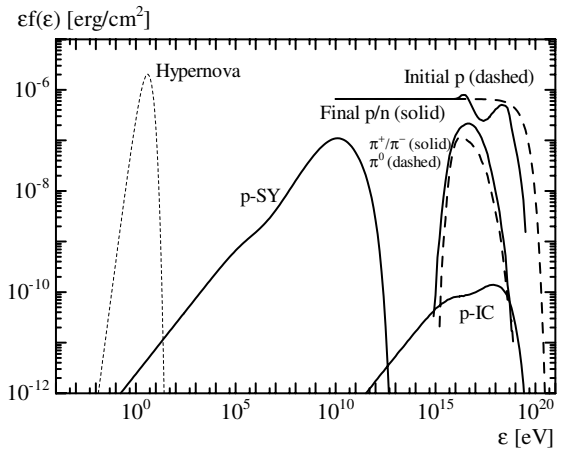}
\caption{
The  primary and secondary particle and photon spectra, similar to Fig. 
\ref{fig:cr-a1}, for a hypernova at the same distance $D=100$ Mpc 
but in a denser wind environemnt of $A_{*}=5$.
\label{fig:cr-a5}}
\end{figure}

Basically, the cooling time-scale is shorter than the integration 
time scale ($\sim 4$ days)
assumed here, which results in a bumpy non-power law 
energy distribution of the  final protons (Fig. \ref{fig:cr-a5}).
The highest energy protons cool via SY
($t_{\rm p \gamma}/t_{\rm p,syn} \sim 1$), while protons 
of $\lesssim 10^{18}$ eV cool via photomeson production
($f_{\rm p \gamma}\sim 0.2$).
If we take into account the radio emission from accelerated electrons,
these complex feature of the proton spectrum may be weakened
because of the high efficiency of pion production above $10^{18}$ eV.

%\vspace*{2cm}

\begin{figure}[h]
\centering
\epsscale{1.0}
\plotone{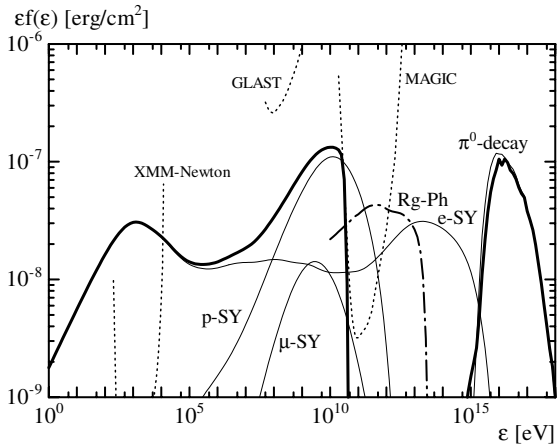}
\caption{
The secondary and tertiary photon spectra, similar to Fig.  \ref{fig:phot-a1} 
but for the hypernova of Fig. \ref{fig:cr-a5} at $D=100$ Mpc in a wind of 
$A_{*}=5$. In addition, the muon synchrotron component (labeled $\mu$-SY),
whose typical energy is determined by a balance between the cooling timescale
and lifetime \citep[see e.g.][]{asa05},
is also plotted.
The dash-dotted line is the regenerated photon spectrum
due to CMB attenuation of $\sim 10^{16}$ eV photons.
\label{fig:phot-a5}}
\end{figure}

In this case the proton SY emission becomes prominent, since
the cooling timescale $\propto \epsilon_{B,-1}^{-3/2} A_*^{-7/2}$ is shorter.
The secondary photon flux (see fig. \ref{fig:phot-a5})
is, as expected, larger than in the lower density wind 
(fig. \ref{fig:phot-a1}) case. However, even this higher density
wind case gives, from D=100 Mpc, an insufficient flux to be detectable 
with GLAST (see Fig. \ref{fig:phot-a5}). However, photons from similar
HNe within 40 Mpc would be detectable by GLAST.
In addition, the 
regenerated TeV photons are promising targets of ACTs.
In Fig. \ref{fig:phot-a5} we plot the regenerated photon spectrum
obtained by the same numerical simulation as in \citet{mur07}.
They are well above the detection limit of present-day ACTs
even for $D=100$ Mpc.

%\vspace*{2cm}

Secondary X-ray photons are emitted by electron-positron pairs
originating from photons due to both proton SY and
SY photons from muon-decay positrons.
This is because the energy of the proton SY photons
shifts higher $\propto \epsilon_{B,-1}^{3/2} A_*^{5/2}$.
Since the energy range of the absorbed proton SY photons
($\sim 10^{11}$ eV)
is narrow, the secondary pairs produce a characteristic peak in 
the fluence spectrum around $0.4 \epsilon_{B,-1}^{1/2} A_*^{3/2}$ keV.
These photons are indirect evidence of proton SY,
and can be easily detected with present-day instruments.

\section{Summary and Discussion}
\label{sec:disc}

We have shown that secondary gamma and X-rays, correlated with the initial
thermal optical emission of hypernovae in the first $\sim 10$ days, can 
provide evidence for proton acceleration, as well as provide a diagnostic 
for amplification  of magnetic field in the blast wave, and for the mass
loss rate in the progenitor stellar wind prior to the explosion.
There are three main spectral components of secondary photons:
1) X-ray photons emitted by electron-positron pairs
originating from $\gamma \gamma$ interactions initiated by
synchrotron photons from muon-decay positrons or protons,
2) synchrotron photons emitted by protons in the GeV band, and
3) regenerated TeV photons emitted by electron-positron pairs
generated by CMB attenuation of $\pi^0$ decay photons
around $10^{16}$ eV.
The X-ray photons are the most promising targets, so that follow-up 
observations of HNe with X-ray telescopes are indispensable to 
find evidence of proton acceleration.  Soft SY photons from 
accelerated electrons may also enhance the electromagnetic cascades 
by interacting with photons from pion-decay.  The interesting $\pi^0$-decay
photon signature (component 3) is also an interesting candidate
for detection with ACTs in the dense wind case ($A_*=5$).

If a HN occurs in our Galaxy at a distance of 10 kpc 
(the rate  for which should be $\siml 10^{-3}-10^{-4}$ yr$^{-1}$),
our results indicate an expected flux $10^{-7}$-$10^{-5}$ erg 
$\mbox{cm}^{-2}$ $\mbox{s}^{-1}$ at $10$ GeV for $A_*=1$-$5$, due 
to proton SY and/or secondary leptons, detectable by GLAST.  
By comparison, the most luminous ``normal" SNRs observed with EGRET 
\citep{esp96} have fluxes of $\sim 10^{-10}$ erg $\mbox{cm}^{-2}$ 
$\mbox{s}^{-1}$ at 10 GeV.  TeV photon detections are not expected 
to be detectable from a Galactic HN, since the photon 
regeneration process mean free path is too long to be effective here.
Our simulations show also that the secondary $\simg$ TeV neutrinos from 
the cascades in a Galactic HN have a spectral peak at $10^{16}$ eV 
with a flux of $10^{-4}$-$10^{-3}$ erg $\mbox{cm}^{-2}$, well above the 
detection limit of IceCube. Thus, one would expect to detect continuous 
TeV neutrino emission for a few days from such Galactic HNe.

For HNe at distances $D\sim 100$ Mpc, the X-rays
(component 1) will be easily detectable by XMM, the sub-TeV radiation
is marginally detectable near the low energy threshold by MAGIC and 
similar ACTs, and the GeV photons from a proton SY (component 2) 
are difficult to detect with GLAST. However, for $A_*=5$,
HNe in the Virgo Cluster 
($D \sim 20$ Mpc) would be easily detectable at GeV energies by GLAST, as 
would also the regenerated TeV photons. Such detections would provide 
constraints on HN models. E.g.  the duration of the proton SY 
emission or the X-ray spectral peak at 1 keV gives an estimate of the 
survival timescale of magnetic fields amplified by the non-linear MHD 
turbulance excited by cosmic rays \citep{bel01}.  Since the maximum proton
energy is  not so sensitive to $\epsilon_B$, the spectral component 1 
due to the cascades from pion production will not change drastically, 
even for $\epsilon_B \ll 0.1$, although the direct proton SY 
(component 2) can become negligible.  In such cases, the disappearance of
the X-ray spectral peak shown in the $A_*=5$ case would be a diagnostic for
such low $\epsilon_B$ values.

The intensity of the cascades depends on the wind density, providing 
a diagnostic for the progenitor mass loss rate. For example, with $A_{*}=0.2$ 
and other parameters as in Fig. \ref{fig:phot-a1} the fluxes are undetectable 
even by X-ray instruments, unless the source is extremely near.
On the other hand, larger values of $A_\ast\sim$ 10 (e.g. as suggested by 
Campana et al (2006) for SN2006aj) would give higher fluxes than those of 
Fig. \ref{fig:phot-a5}, enhancing the probability of detection at $D\sim
100$ Mpc at TeV, GeV, and X-ray energies.

\begin{acknowledgments}
We thank K. Nakazawa and K. Murase for providing the sensitivity curves
of various instruments, S. Razzaque and S. Inoue for valuable discussions, 
the referee for valuable comments, and NSF AST 0307376 for partial support.
\end{acknowledgments}

%\clearpage

\end{document}